# ANALYSIS OF THE USER ACCEPTANCE FOR IMPLEMENTING ISO/IEC 27001:2005 IN TURKISH PUBLIC ORGANIZATIONS

Tolga MATARACIOGLU[1] and Sevgi OZKAN[2]

[1]TUBITAK National Research Institute of Electronics and Cryptology (UEKAE),
Department of Information Systems Security, 06700, Ankara, TURKEY
[2]Middle East Technical University, Informatics Institute,
Department of Information Systems, 06531, Ankara, TURKEY

mataracioglu@uekae.tubitak.gov.tr, sozkan@ii.metu.edu.tr

*ABSTRACT*

*This study aims to develop a model for the user acceptance for implementing the information security standard (i.e. ISO 27001) in Turkish public organizations. The results of the surveys performed in Turkey reveal that the legislation on information security public which organizations have to obey is significantly related with the user acceptance during ISO 27001 implementation process. The fundamental components of our user acceptance model are perceived usefulness, attitude towards use, social norms, and performance expectancy.*

## I. INTRODUCTION

Information Security Management System (ISMS) is a set of processes and the main goal of those systems is to manage information security issues in an enterprise [36]. ISMS uses Plan-Do-Check-Act (PDCA) model and the input of this model is information security requirements and expectations. The expected output is obviously managed information security. In Plan phase, establishing ISMS policy, objectives, processes and procedures relevant to managing risk and improving information security to deliver results in accordance with an organization's overall policies and objectives are performed. In Do phase, implementation and operation of the ISMS policy, controls, processes and procedures are performed. In Check phase, assessment and, measurement of process performance against ISMS policy, objectives and practical experience and reporting of the results to management for review are performed. In Act phase, taking corrective and preventive actions, based on the results of the internal ISMS audit and management review or other relevant information, to achieve continual improvement of the ISMS is performed.

ISO 27001 is the standard of implementing ISMS to an enterprise. To implement each of the phase of the PDCA model, items numbered 4.2.1, 4.2.2, 4.2.3 and 4.2.4 are given in the standard. The standard has ten domains including security policy, organizational security, asset classification and control, access control, compliance, personnel security, physical and enviromental security, system development and maintenance, communications and operations management and business continuity management.

Based on the theory of reasoned Action, Davis [37] developed the Technology Acceptance Model (TAM) which deals more specifically with the prediction of the acceptability of an information system. This is an information system, the probability that he will use it is high if he perceives that the system will improve his performance at work. Besides, the Technology Acceptance Model hypothesizes a direct link between perceived usefulness and perceived ease of use. With two systems offering the same features, a user will find more useful the one that he finds easier to use.

The UTAUT (Unified Theory of Acceptance and Use of Technology) aims to explain user intentions to use an IS and subsequent usage behavior. The theory holds that four key constructs (performance expectancy, effort expectancy, social influence, and facilitating conditions) are direct determinants of usage intention and behaviour. Gender, age, experience, and voluntariness of use are posited to mediate the impact of the four key constructs on usage intention and behavior. The theory was developed through a review and consolidation of the constructs of eight models that earlier research had employed to explain IS usage behaviour (theory of reasoned action, technology acceptance model, motivational model, theory of planned behavior, a combined theory of planned behavior/technology acceptance model, model of PC utilization, innovation diffusion theory, and social cognitive theory). Subsequent validation of UTAUT in a longitudinal study found it to account for 70% of the variance in usage intention.

## II. LITERATURE REVIEW

We will perform a literature review about user acceptance in this section concerning the last 10 years' journals.

[1] says that important organizational barriers hinder the implementation of personalized e-government services and important user obstacles, such as access, trust, control, and privacy, have to be overcome to make fruitful use of those personalized e-government services.

[2] considers the following questions: What are the limitations of the existing TAM for studying virtual community? What is effect of social networks on user acceptance of technology for virtual community? and How can the influence of different types of social ties serve as a basis for exploring the user acceptance of technology in a virtual community? They explore the possibility for extending TAM to incorporate the influence of the different types of social ties as a new theoretical construct.

[3] attempts to provide a systematic analysis of the explanatory and situational limitations of existing technology acceptance studies. Ten moderating factors are identified and categorized into three groups: organizational factors, technological factors and individual factors.

[4] proposed a revised technology acceptance model for measuring end user computing (EUC) acceptance. An empirical study was conducted to collect data. This data was empirically used to test the proposed research model. The structural equation modeling technique was used to evaluate the causal model and confirmatory factor analysis was performed to examine the reliability and validity of the measurement model.

A new policy approach is proposed to increase ICT acceptance in [5]. The approach is based on strategies of segmentation and differentiation. This entails that policy initiatives are specifically targeted towards different groups in the population.

[6] investigated four potential moderating variables – perceived risk, technology type, user experience, and gender – in users' technology adoption. Their moderating effects were tested in an empirical study of 161 subjects. Results showed that perceived risk, technology type, and gender were significant moderating variables.

Using the theory of planned behavior (TPB) as a theoretical framework, [7] investigates the effect of a set of antecedent factors on the intention to accept EDMS. Collected from a sample of 186 users of real e-Government's electronic document management systems (EDMS) in Taiwan, the results strongly support the utilization of TPB in predicting users' intention to accept EDMS. In addition, the findings indicate that perceived usefulness, perceived ease of use, training, compatibility, external influence, interpersonal influence, self-efficacy, and facilitating conditions are significant predictors of users' intention to utilize EDMS.

Analysis of the data in [8] showed that users and analysts did not agree on the user's involvement nor did they agree on their perceptions of the acceptability of the system to the user. Relationships of self-ratings of user involvement (UI) with system usage and system acceptance by the user demonstrated high correlations, which were attributed to the narrow focus of the UI and system acceptance measures rather than the original more global measure.

[9] tested one of TAM's assumptions that the 'perceived ease-of-use' and 'perceived usefulness' constructs fully mediate the influence of external variables on usage behaviors. Using a survey of 125 employees of a U.S. Government agency they found, contrary to the normally accepted assumption, that external variables could have direct effects on usage behavior over and above their indirect effects.

[10] examines the levels of security acceptance that can exist amongst employees within an organization, and how these levels relate to three recognised levels of corporate culture. It then proceeds to identify several factors that could be relevant to the development of culture, from traditional awareness-raising techniques through to context-aware promotion of security.

[11] proposed an extended model for analyzing initial IT acceptance behavior of Chinese users. To empirically test the model, they conducted a survey regarding the recognition and adoption of an English e-learning system in the freshmen of a business school.

Using a survey sample collected from 722 knowledge workers using desktop computer applications on a voluntary basis in Saudi Arabia, [12] examined the relative power of a modified version of UTAUT in determining 'intention to use' and 'usage behavior'. They found that the model explained 39.1% of intention to use variance, and 42.1% of usage variance.

A statistical meta-analysis of the technology acceptance model (TAM) as applied in various fields was conducted using 88 published studies that provided sufficient data to be credible in [13]. The results show TAM to be a valid and robust model that has been widely used, but which potentially has wider applicability.

Relying on social exchange theory and technology acceptance literature, [14] argues that users' acceptance of a newly introduced technology is related to the beliefs that individuals have about their social relationships with the supervisor, the team members, and the organization as a whole. Besides confirming previous findings in the acceptance domain, their results demonstrate that perceived usefulness and ease of use about technology are influenced by team–member exchange, leader–member exchange, and organizational support, while affective commitment toward the organization shows a positive influence only on perceived usefulness.

In [15], an attempt is made to explain the descriptive data of a large-scale representative survey of the use of government Internet services by the Dutch population in 2006 by means of a multidisciplinary model of technology acceptance and use that is applied to these services.

[16] concludes that TAM is a useful model, but has to be integrated into a broader one which would include variables related to both human and social change processes, and to the adoption of the innovation model.

[17] describes the development of a comprehensive model for measuring user satisfaction in the context of E-Government. It rethinks the e-strategies of government and subsequently presents a conceptual model derived from ICT acceptance theory. Both quantitative as well as qualitative research have been carried out in order to elaborate the model and to formulate adequate indicators for measuring user satisfaction.

Attitude theory from psychology provides the rationale for hypothesized model relationships, and validated measures were used to operationalize model variables in [18]. A field study of 112 users regarding two end-user systems was conducted to test the hypothesized model. TAM fully mediated the effects of system characteristics on usage behavior, accounting for 36% of the variance in usage.

[19] develops hypotheses about the effects of the dimensions (innova-tiveness, optimism, discomfort, and insecurity) of technology readiness on two key stages of Internet acceptance, adoption, and usage of different Internet-based activities, and test them through a two-stage model using U.S. consumer survey data.

[20] reported the results of an investigation that compared two versions of technology acceptance model (TAM) in understanding the determinants of user intention to use wireless technology in the workplace. The first model is derived from original TAM that includes perceived usefulness, perceived ease of use, attitude and behavioral intention. The alternative model is a parsimonious version in which the attitude was taken out as was done by many prior studies.

[21] explores the extent of user acceptance of the national identity card (NIC) and driving license (DL) applications embedded in the Malaysian government multipurpose smartcard (called MyKad).

[22] assessed the use of "Personal Computer Technology" in public organizations of developing countries in South Asia, particularly in Pakistan.

[23] considers whether people likely to pursue careers within the not-for-profit (NFP) sector have different attitudes to technology and whether such differences affect the measures used within technology acceptance models.

The model is tested using data gathered from 374 end users of the Internet in Korean firms and data analysis is conducted using a structural equation modeling with LISREL in [24]. Significant relationships are found between experience and usefulness, between experience and ease of use, and between ease of use and usefulness.

[25] says that, some drawbacks must be considered. Important organizational barriers hinder the implementation of personalized e-government services and important user obstacles, such as access, trust, control, and privacy, have to be overcome to make fruitful use of those personalized e-government services.

Survey of nonprofit public relations practitioners (N = 409) applied the Unified Theory of Acceptance and Use of Technology (UTAUT) in [26]. Findings indicate that women consider social media to be beneficial, whereas men exhibit more confidence in actively utilizing social media.

[27] examined the influence of information technology (IT) acceptance on organizational agility. The study was based on a well-established theoretical model, the Technology Acceptance Model (TAM). They attempted to identify

the relationships between IT acceptance and organizational agility in order to see how the acceptance of technology contributes to a firm's ability to be an agile competitor. Structural equation modeling techniques were used to analyze the data.

[28] argues that the frequent organizational incoherence and thus the unviability of modern technology arises from 'social alienation' between the innovation-commitment phase and the implementation of the technology in society. The roles of technical elites and of particular concepts of technology in this alienation are emphasized.

The findings show in [29] that the majority of Jordanian public organizations have access to computers, and most of the agencies have in-house systems. Microcomputers were found to be the most common computer platform, and agency use of IT was consistently intense in the area of financial management. Most public organizations with in-house computers employ on-staff programmers, whose numbers range from one to four or fewer.

[30] examined the efficacy of this psychologically based TAM within two samples of government workers experienced in the use of computers (N=108). All participants completed a self-report questionnaire consisting of both previously developed and purpose derived scales. The study achieved its purpose of replicating and validating a development of the TAM, although only moderate support for the model was found.

[31] develops hypotheses about the effects of the dimensions (innovativeness, optimism, discomfort, and insecurity) of technology readiness on two key stages of Internet acceptance, adoption, and usage of different Internet-based activities, and test them through a two-stage model using U.S. consumer survey data.

[32] examines the diffusion process and those factors which determine whether or not an innovation is accepted with special note towards the determinants of customer acceptance of an innovation.

Based on the revised unified theory of acceptance and use of technology (UTAUT) model, [33] described a theoretical framework that incorporates the unique characteristics of m-commerce to enhance our understanding of m-commerce acceptance and usage in China.

Thirty-four articles selected from international journals were analyzed in [34] to show that most of the relationships in the classical TAM are significant, but the stabilities of these relationships differ. The significant positive relationships between perceived ease of use and its independent variables are more stable than the others. Various factors can strengthen or weaken these relationships.

[35] examines the evidence that the TAM predicts actual usage using both subjective and objective measures of actual usage. They performed a systematic literature review based on a search of six digital libraries, along with vote-counting meta-analysis to analyze the overall results. The search identified 79 relevant empirical studies in 73 articles. The results show that BI is likely to be correlated with actual usage. However, the TAM variables perceived ease of use (PEU) and perceived usefulness (PU) are less likely to be correlated with actual usage.

## III. MODEL AND METHODOLOGY

We have used surveying as our methodology and asked people questions about general, perceived usefulness, attitude towards use, performance expectancy and social norms.

Our new model for user acceptance for implementing ISO/IEC 27001:2005 in Turkish public organizations is given in Fig 1.

Our questions included in the survey are given below with a classification by general, perceived usefulness, attitude towards use, social norms, and performance expectancy.

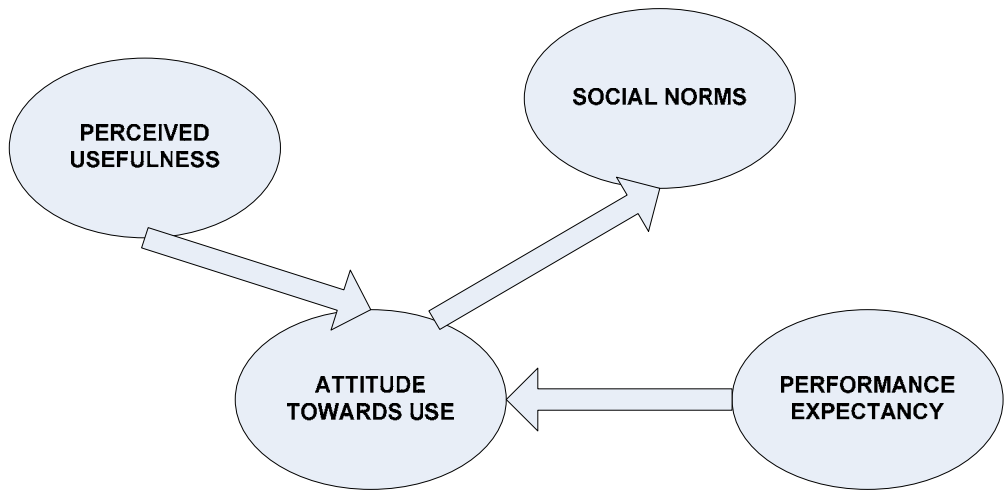

Fig 1. The new model for the user acceptance of ISO 27001

*General:*

GN1) Gender?
   a) Female
   b) Male
GN2) Age?
   a) Under 20
   b) Between 20 and 30
   c) Between 31 and 45
   d) Above 45
GN3) Academic level?
   a) High school
   b) Bachelor
   c) Master
   d) PhD
   e) Other
GN4) Description of organization?
   a) Public organization
   b) Private sector
   c) Armed Forces
   d) I do not work
GN5) Why do you use information systems?
   a) For business
   b) To read my e-mails
   c) For internet surfing
   d) I do not use information systems
GN6) How much time do you spend with information systems?
   a) Approximately 1 hour
   b) Less than 1 hour
   c) Between 1 and 3 hours
   d) More than 3 hours
GN7) What is your privilege in your computer?
   a) Administrator
   b) User
   c) I do not use computer
GN8) Does your organization have the ISO 27001 certificate?

   a) Yes
   b) No

*Perceived Usefulness:*

PU1) ISO 27001 will improve my quality of work
   a) Strongly agree
   b) Agree
   c) I'm not sure
   d) Disagree
   e) Strongly disagree
PU2) ISO 27001 will make it easier to do my job.
   a) Strongly agree
   b) Agree
   c) I'm not sure
   d) Disagree
   e) Strongly disagree
PU3) ISO 27001 will make it faster to do my job.
   a) Strongly agree
   b) Agree
   c) I'm not sure
   d) Disagree
   e) Strongly disagree
PU4) ISO 27001 will give me better control over my job.
   a) Strongly agree
   b) Agree
   c) I'm not sure
   d) Disagree
   e) Strongly disagree
PU5) ISO 27001 will enhance my effectiveness.
   a) Strongly agree
   b) Agree
   c) I'm not sure
   d) Disagree
   e) Strongly disagree
PU6) ISO 27001 will make it more secure to do my job.
   a) Strongly agree
   b) Agree
   c) I'm not sure
   d) Disagree
   e) Strongly disagree

*Attitude Towards Use:*

AU1) I think that ISO 27001 certification is good for my organization.
   a) Strongly agree
   b) Agree
   c) I'm not sure
   d) Disagree
   e) Strongly disagree
AU2) I think that ISO 27001 certification is beneficial for my organization.
   a) Strongly agree
   b) Agree
   c) I'm not sure

   d) Disagree
   e) Strongly disagree
AU3) I think that ISO 27001 certification is garbage for my organization.
   a) Strongly agree
   b) Agree
   c) I'm not sure
   d) Disagree
   e) Strongly disagree
AU4) I think that ISO 27001 certification is efficient for my organization.
   a) Strongly agree
   b) Agree
   c) I'm not sure
   d) Disagree
   e) Strongly disagree
AU5) I think that ISO 27001 certification is unnecessary for my organization.
   a) Strongly agree
   b) Agree
   c) I'm not sure
   d) Disagree
   e) Strongly disagree

*Social Norms:*

SN1) Which one should be implemented first so as to establish information security management system in your organization?
   a) COBIT
   b) ISO 27001
   c) Legal infrastructure
   d) All of above
   e) None
SN2) Do you want that ISO 27001 certification is a must in our country?
   a) Strongly agree
   b) Agree
   c) I'm not sure
   d) Disagree
   e) Strongly disagree
SN3) Who do you think is responsible from obtaining ISO 27001 certificate and establishing information security management system?
   a) Senior management
   b) IT
   c) Consultants
   d) All organization employees
   e) I do not know
SN4) Suppose that ISO 27001 certification is a must in our country. Which one is true for you?
   a) I meet all of the requirements of ISO 27001 on paper since it is a must, however my method in doing my job remains same as before.
   b) I do my job concerning that both ISO 27001 is a must and it brings effectiveness to my job.
   c) ISO 27001 is not applied in my organization even on paper assuming that it is a must in our country.
SN5) There should firstly be made laws so as to establish information security management system in my organization. Otherwise, my organization will not take notice this certification.
   a) Strongly agree
   b) Agree
   c) I'm not sure
   d) Disagree

e) Strongly disagree

*Performance Expectancy:*

PE1) I think that there will be a decrease in my performance when ISO 27001 certification comes to my organization.
   a) Strongly agree
   b) Agree
   c) I'm not sure
   d) Disagree
   e) Strongly disagree

PE2) I think that there will be an increase in plodding jobs when ISO 27001 certification comes to my organization.
   a) Strongly agree
   b) Agree
   c) I'm not sure
   d) Disagree
   e) Strongly disagree

PE3) I think both job doing performance and ISO 27001 can be driven together.
   a) Strongly agree
   b) Agree
   c) I'm not sure
   d) Disagree
   e) Strongly disagree

PE4) I think both job doing rate and ISO 27001 can be driven together.
   a) Strongly agree
   b) Agree
   c) I'm not sure
   d) Disagree
   e) Strongly disagree

PE5) I think both job doing efficiency and ISO 27001 can be driven together.
   a) Strongly agree
   b) Agree
   c) I'm not sure
   d) Disagree
   e) Strongly disagree

## IV. DISCUSSION

We determined our model for user acceptance of ISO 27001 consisting of four components as perceived usefulness, attitude towards use, social norms, and performance expectancy. Also we expect that performance expectancy and perceived usefulness affect attitude towards use. Further, all of those three components affect social norms. We conclude that the legislation should be enacted after adopting ISO 27001 in organizations. Otherwise, organizations will behave ISO 27001 as a plodding workload.

As a result, user adoption in ISO 27001certification in Turkish public organizations is significantly related with legislation and the precedence should be given to user acceptance when comparing legislation enacting with user acceptance.

## V. CONCLUSIONS

In this paper, we tried to analyze of the user acceptance for implementing ISO/IEC 27001:2005 in Turkish public organizations. First, we introduced the concepts ISMS, PDCA, ISO 27001, user acceptance, TAM, and UTAUT in Part I. Then in Part II, we presented our new model and the questions of our survey which determined our model components. In Part IV, we analyzed the survey results and discussed the results in Part V.

The results of the surveys performed in Turkey reveal that the legislation on information security public which organizations have to obey is significantly related with the user acceptance during ISO 27001 implementation process. The fundamental components of our user acceptance model are perceived usefulness, attitude towards use, social norms, and performance expectancy.

**Tolga MATARACIOGLU**

After receiving his BSc degree in Electronics and Communications Engineering from Istanbul Technical University in 2002 with high honors, he received his MSc degree in Telecommunications Engineering from the same university in 2006. He is now pursuing his PhD degree in Information Systems from Middle East Technical University. He is working for TUBITAK National Research Institute of Electronics and Cryptology (UEKAE) at the Department of Information Systems Security as senior researcher. He is the author of many papers about information security published nationally and internationally. He also trains various organizations about information security. His areas of specialization are system design and security, operating systems security, and social engineering.

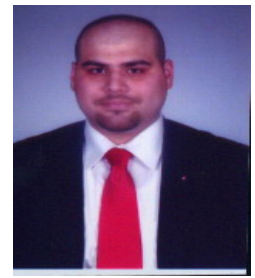

**Sevgi OZKAN**

Dr Sevgi Ozkan is an Assistant Professor at Department of Information Systems, Informatics Institute, Middle East Technical University Turkey. She is currently the Associate Dean of the School. She holds a BA and an MA in Engineering from Cambridge University and an MSc in Business Information Systems London University UK. She has a PhD in Information Systems Evaluation. She is a Fellow of the UK Higher Education and a Research Fellow of Brunel University UK. Since 2006, Dr. Ozkan has been involved with a number of EU 7th Framework and National projects in e-government.

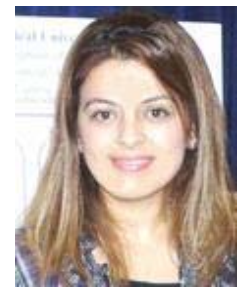